\documentstyle [12pt,amssymb,english,epsf] {article}
\begin{document}

\begin {flushright}
{30-99}
\end{flushright}
\vspace {2.0cm}

\begin{center}
{\bf{\large{QUASI-COHERENT INTERACTION OF PROTONS  WITH \\
$^{28}Si$ AT $T_p$=1 GEV}}}

\vspace {2.0cm}

Institute of Theoretical and Experimental Physics, ITEP,\\
117259 Moscow, Russia\\
\vspace {2.5cm}

M.P.Besuglov, B.M.Bobchenko, E.V.Bustritskaya, A.A.Vasenko,
\fbox{M.E.Vishnevsky},
N.D.Galanina, K.E.Gusev, V.S.Demidov, E.V.Demidova, V.V.Zchurkin,
I.V.Kirpichnikov, V.A.Kuznetsov,  V.N.Markizov, M.A.Martemianov,
A.A.Nedosekin, B.N.Pavlov,\fbox{V.A.Sadyukov},  A.Yu.Sokolov,
A.S.Starostin, N.A.Khaldeeva
\end{center}

\begin{center}
\vspace{1.3cm}
ABSTRACT
\vspace {1.0cm}
\end{center}
\begin{quotation}

The total, elastic and inelastic cross-sections for quasi-coherent interactions
of protons with $^{28}Si$ have been measured using hadron-gamma coincidence
method at the energy 1 GeV. The limits for the existence of long-lived excited
states of the $^{28}Si$ nucleus have been obtained in the range of energies
from 0 to 0.8 GeV.

\end{quotation}

\clearpage

\section{Introduction}

Quasi-coherent hadron-nucleus interactions $A(h,h^{\prime})A^\ast$
which result in nucleus $A$ excitation and hadron $h$ either elastic
scattered or transformed into a system of hadrons can provide information
on the properties of nucleus as well as on the mechanism of hadron-nuclei
interaction.  The experimental study of quasi-coherent reactions was
performed mostly at high accuracy magnetic spectrometers measuring
the scattering angle and nucleus momentum transfer. Reactions with the
excitation of certain levels were selected via the missing mass.
The investigation was carried out in a narrow region of transferred energies,
since at the excitation energies over $\approx$ 10 MeV the domination of
nucleus disintegration processes does not allow to separate quasi-coherent
reactions without nuclear states being fixed.

In this paper quasi-coherent reactions are investigated using the method of
hadron-gamma coincidence based on a simultaneous registration of the leading
hadron and instant $\gamma$-rays following the transition of an excited
nucleus into the state with minimal excitation energy or into
the ground state. The identification of reactions is carried out via the
energy of registered $\gamma$. The kinematical parameters measured by the
magnetic spectrometer are used for the analysis of hadron-nucleus reactions.

Such method does not limit the value of energy transferred by the scattering
nucleus.  The region of possible energy transfer include
the region of elastic hadron interactions as well as
the region where the production of $\pi$-mesons
with total charge equal to zero is allowed by kinematic laws.

It is evident that in the first region the total amount of transferred
energy is spend for the nucleus excitation. For the hadron energy about
1 GeV the excitation of three lowest levels with natural parity
has been observed in LINP experiment (~\cite{Alhaz3}~--~\cite{Alhaz5}).
These data were used later to obtain the distribution of the density of
nuclear matter (~\cite{Alhaz1}, ~\cite{Alhaz2}).
Slightly above the threshold of nucleon production gigantic resonances
can be found. Their excitation in quasi-coherent reactions is not yet
investigated well enough although such experimental data would be important
for the estimation of radiative decay probability of one- and multiphonon
gigantic nuclear resonances. The first experimental evidence of the radiative
decay of two-phonon resonances was obtained recently in nucleus-nucleus
interactions ~\cite{Chomaz}.  In the same energy region the narrow resonance
states  discovered in 1982 and interpreted as the cluster oscillations
in nuclear matter can be displayed (~\cite{Wuos}--~\cite{Bauh}).

In the region of higher transferred energies where non-elastic processes of
meson production are possible quasi-coherent reactions have not been
extensively studied.
The abnormal contribution of non-elastic processes with extra particle
production was observed by the method of hadron-gamma coincidence
for the reactions  $^{16}O(\pi,\pi X)^{16}O^*$ and $^{40}Ca(\pi,\pi X)
^{40}Ca^*$ ( ~\cite{Kirp1}, ~\cite{Kirp2}, ~\cite{Kirp3}).
Quasi-coherent $\pi$-meson production has been investigated
only for the ($\pi$,Si) reaction  \cite{pic3pi},
although a number of theoretical works indicate that for the understanding of
the hadron-nuclei interaction mechanism other reactions must be studied.
 ~\cite{Korot}.
The existence of nuclei states with high excitation (180 MeV)
has been assumed in ~\cite{Kosov} in order to explain
the maximum in the energy transfer distribution in the charge-exchange
reaction $A(\pi,\pi\pi)X$.

In the present work the quasi-coherent interaction $^{28}Si(p,px)^{28}Si^*$ is
investigated using the method of hadron-gamma coincidence in the region of
transferred energies from 0 to 0.8 GeV in order to discover long-lived and
highly excited nuclei levels with electromagnetic decay channels
and to study the mechanism of meson production.

\section{The experiment}

The experiment was carried out at the ITEP proton synchrotron using MAGE
(MAgnetic-GErmanium spectrometer).

\subsection{\hspace{-0.8em}. The beam}

The proton beam was formed by the magnetic tract consisting of 3 rejecting
magnets and 4 focusing magnetic lenses placed at the angle $3.5^\circ$ to the
direction of inner accelerator beam.
Protons accelerated to the momentum of 1.83 GeV/c in the synchrotron
scattered on inner Be target and falled into the magnetic tract.
The beam intensity was $10^{6}$ protons/sec per $\approx$ 0.4 sec.
The momentum distribution of protons reaching the experimental target had
a gaussian shape with the mean value of 1.822$\pm$0.001 GeV/c and the standard
deviation of 0.026 GeV/c.
The mean value depends on the beam tract tuning and can vary withing the
limits of 0.015 GeV/c for different runs (the correspondent corrections
were included in data analysis). The angle beam dispersion in the target
area did not exceed $1.2^{0}$, the size of the beam spot was $4\times4~cm^2$.

\subsection{\hspace{-0.8em}. The set-up components}

A schematic view of the MAGE spectrometer is presented in fig.1.

\begin{figure}[hbt]
\begin{center}
\mbox{ \epsfysize=7.5cm  \epsffile{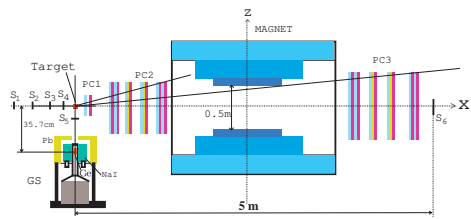} }
\end{center}

\caption{ The side view of the MAGE spectrometer.}

\label{magpic1}

\end{figure}

The MAGE spectrometer is a two-arm device. One arm is a
magnetic spectrometer with proportional chambers, the second is a
gamma-spectrometer based on a Ge(Li)- detector surrounded by a
protection shield of NaI scintillation counters which is
connected in anticoincidence with the detector.

\subsubsection{\hspace{-1em}. The magnetic spectrometer}

The magnetic spectrometer includes the magnet, track detectors and a system of
scintillation counters. The working area of the magnetic field is
$0.5\times1\times1.8m^{3}$ with the field value in the center
equal to 0.72 Tl.

The tracking part of the spectrometer consists of 30 plane multiwire
proportional chambers ~\cite{Ale86} arranged in 3 groups: PC1, PC2
(placed between the target and the magnet) and PC3 (placed behind
the magnet). The group PC1 includes 2 rectangular $240\times240~mm^{2}$
coordinate planes with wires oriented transverse to the beam direction.
The group PC2 includes 14 $480\times1280~mm^{2}$ planes: 6 with wires placed
along the beam direction (z), 6 with transverse wires (y), in 2 planes
wires are placed at the angle $\approx37^{0}$ with respect to the beam
direction (w). The group PC3 used for the momentum measurements has the same
structure as PC2 but the chamber size is $880\times1840~mm^{2}$.
The distance between the signal wires is 2.5 mm, in slope planes - 2 mm,
the total number of wires - about 15 000.

The proportional chambers are filled with the gas mixture
$Ar~(70\%)~+~CO_{2}~(29,8\%)~+CF_{3}Br~(0,2\%)$.

The minimal measured momentum is 0.4 GeV/c, the precision of momentum
measurement is $\Delta p/p(\%)= ~0.32p(\mbox{GeV/c})+0.57$.

The target was placed 2.56 m from the center of the magnet.
The maximum deviation angle of secondary charged particles registered
by the spectrometer did not exceed 0.25 rad for the momentum measurements
and increased to 0.55 rad when only PC1 and PC2 have been used.

The direction of charged secondary particles was measured with the precision
of 0.002 rad, while the scattering angles with respect to
the beam direction were determined with less precision (0.008 rad) because of
the beam angle dispersion.

The system of scintillation counters $S_{1} - S_{6}$ was used to select
the beam spot with the size of the target and to generate a signal of
interaction of beam particles with the target. Together with the
signal of $\gamma$-detector it is a part of the master signal triggering
on-line data acquisition system based on VAX4200 connected with other
computers by ETHERNET.  The read-out electronics is described in
~\cite{Bogd86} and ~\cite{Men87}.
The maximum rate of data acquisition is about 400 kb/sec.

\subsubsection{\hspace{-1em}. Germanium $\gamma$-spectrometer }

The Ge(li)-NaI(Tl) $\gamma$-spectrometer (GS in fig.1) has been employed
to register the instant photons following the deexcitation of
atomic nuclei and to measure the $\gamma$-energy ~\cite{Vas97}.

The spectrometer includes Ge(Li) detector with the volume 100 $cm^{2}$  placed
in a cryostat and surrounded by a six-section NaI(Tl)
shaft assembly of 300 mm height, 350 mm outer diameter and 100 mm shaft
diameter connected in anticoincidence with the detector.
Ge detector is placed 37.5 cm from the centre of the target.
The scintillation counter $S_{5}$ working in anticoincidence prevented
the triggering of the spectrometer by charged particles. To reduce the
outside background the spectrometer was surrounded by lead protection.
The analog signal from the Ge detector was digitized by ADC
and transferred to the computer. The monitoring of $\gamma$-spectrometer
operation was performed by the 4096-channel NOKIA analyser.

The energy calibration of the Ge detector was performed using 18
$\gamma$-lines of $^{226}Ra$ in the range of energies from 180 to 3100 KeV.
It was approximated by the function

\begin{equation}
\label{feg}
 E_{\gamma}(\mbox{keV})=b(A_{gd}-a)^{c},
\end{equation}

where $A_{gd}$ is the channel number of the amplitude analyser
or ADC corresponding to the energy $E_{\gamma}$.

The values a=-48.45$\pm$0.07, b= 0.9532$\pm$0.0007,
and c=0.9849$\pm$\\0.0003 were estimated by the method of least squares.

The total $\gamma$-spectrometer efficiency
depends on two factors: $\Omega_{eff}(E_\gamma)=
\Omega_g(E_\gamma)\cdot\eta(E_\gamma)$,
where $\Omega_g(E_\gamma)$ is the geometric efficiency, and
$\eta(E_\gamma)$ is the efficiency of the master signal.

To determine the geometric efficiency $\Omega_g(E_\gamma)$ $\gamma$-spectra
from 4 sources $^{60}Co$, $^{88}Y$, $^{137}Cs$ and $^{228}Th$ with well
known intensities which cover the investigated energy range (0.5-3.0 MeV)
have been measured by the spectrometer.
The geometric efficiency was estimated for 8 experimental points by the
method of least squares as
$ln\Omega_g(\mbox{mster})= (11.02\pm0.05) -
(2.08\pm0.01)lnE_\gamma(\mbox{keV})+(0.0793\pm0.0007)lnE_
\gamma^2(\mbox{keV})$  with the precision less than 2\%.

The introduction of the coefficient $\eta(E_\gamma)$ is connected with the
use of time-gating for the reduction of the background.  For runs with the
target the "gate" duration was limited to $\delta$t=50~nsec.
Due to large fluctuations in
the fore-front of Ge detector signal it was followed by the deterioration of
the spectrometer efficiency. The efficiency $\eta(E_\gamma)$ was obtained by
comparison of two $\gamma$ spectra for $\delta$t=50~nsec and
$\delta$t=250~nsec and approximated by
the function $\eta(E_\gamma)=0.7- 0.55\cdot exp(-0.017E_\gamma(\mbox{keV}))$.

 The dependence of $\Omega_{eff}$ on photon energy is shown in Fig.2 in
units of the "efficient" solid angle $\Omega_{eff}$ covered by an
ideal (detecting each photon) $\gamma$-detector.

\begin{figure}[hbt]

\begin{center}

\mbox{ \epsfysize=9.5cm \epsfxsize=12cm \epsffile{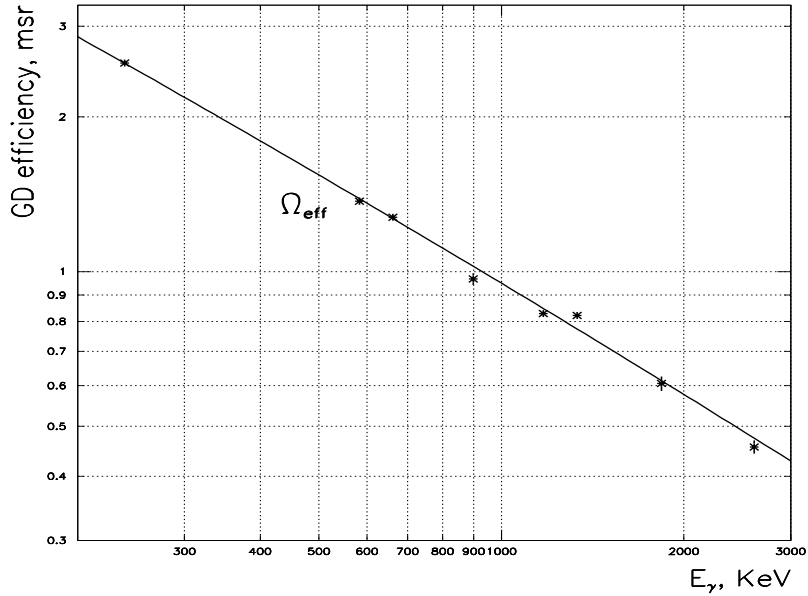} }

\end{center}

\caption{The dependence of Ge spectrometer efficiency on
$\gamma$ energy.}

\label{effelecg}

\end{figure}

The anticoincidence signal from NaI allows to suppress the
background in $\gamma$-spectrum of the Ge detector
by a factor of 20-30 for runs with target (see fig.3).

\begin{figure}[hbt]

\begin{center}

\mbox{ \epsfysize=9.5cm \epsffile{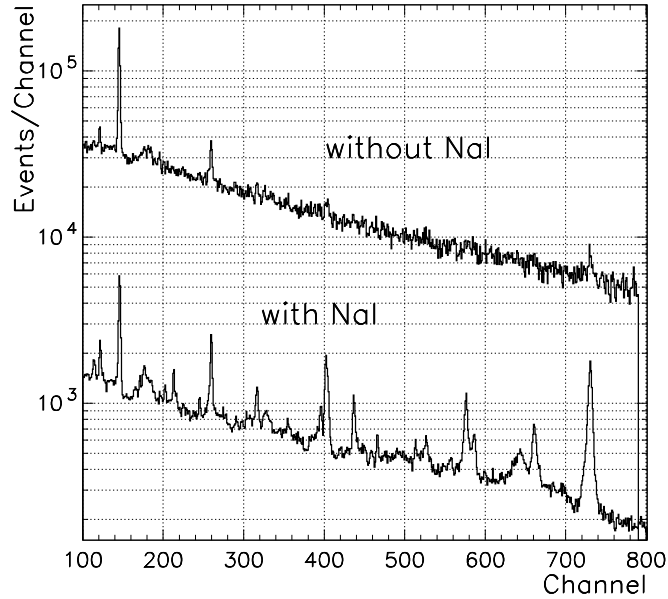} }

\end{center}

\caption{The supression of background continuum by active NaI shield.
Top histogram presents the region of $\gamma$-spectrum without active
NaI protection, bottom - the same region with switched protection.}

\label{gdnai}

\end{figure}

To estimate the systematic errors in cross-section measurements
caused by the uncertainty in $\gamma$-spectrometer efficiency, the procedure
of efficiency definition was repeated several times.  It was found that the
systematic errors do not exceed $\pm$10\% while the basic contribution
comes from the uncertainty in $\eta(E_\gamma)$.

The employment of Ge detector in our experiment has created a major difficulty
which is the presence of background signals from charged particles and
$\gamma$-quanta of higher energies.
The amplitude overload of the spectrometer tract results in uncontrolled loss
of spectrometer efficiency in case of the standard electronics, since the
preamplifier signal has a duration of several milliseconds.

To prevent the overload the charge collection time defined by the
preamplifier feedback circuit has been decreased by two orders.
The energy resolution of Ge-detector (FWHM) was found to be 9-16 keV in
the range of $\gamma$ energies of 0.5-2.0 MeV and the shape of the detector
response has been well approximated by Gaussian function.

 To obtain the calibration dependence the average values for
 $\gamma$-lines of $^{226}Ra$ spectrum have been used.

The limitation of the time delay of a registered signal from Ge detector
with respect to a signal from scattered proton to the value
of 50 nsec imposes an upper limit on the half-decay period $T_{1/2}$ for
the observed levels.  Apparently the intensities of the
levels with $T_{1/2}\le10\div20~nsec$ can be measured without distortion,
for levels with higher $T_{1/2}$ it is necessary to take into account
non-registered decays.

\subsection{\hspace{-0.8em}. Experimental procedure}

Three runs have been carried out for 1.83 GeV/c proton beam
directed at the center of a Si cylinder target with the
diameter of 80 mm and the height of 27 mm.
The isotopic structure of the target was the following: $^{28}Si$--92.23\%,
$^{29}Si$--4.67\%, $^{30}Si$--3.1\%, the density - 2.20 g/cm$^3$.
The interaction of protons with the most widespread isotope $^{28}Si$ has
been studied. For this isotope the quantum number of the ground state is
0$^+$, the majority of higher excited nuclear states discharge through the
first excited level (2$^{+}$) with the energy 1.77003 MeV and half-
decay period (686~$\pm$~13~)$10^{-15}$ sec ~\cite{Endt}.
 The threshold energy of knocking out a nucleon from $^{28}Si$ nucleus
is $E_t$=11.58 MeV.

To select all the events of proton interaction with the
target followed by the detection of a photon in GS a master signal was
generated according to the scheme $S_1S_2S_3S_4\overline{S_5S_6NaI}Ge$.

 The counters $S_1$--$S_4$ limited the beam transverse size
to $4\times4$ cm$^2$.  The counter $\overline{S_6}$,
connected in anticoincidence, rejected particles passing through the target
without interaction in the angle $\le 0.4^\circ$. The counter ${S_5}$
and NaI-assembly provided the suppression of background in $\gamma$-detector.

The master signal triggers the work of the read-out system which transfers
into the computer the addresses of the wires hit by charged particles,
the amplitude $A_{gd}$ of Ge detector signal and the indications of recount
monitoring devices. The on-line information is send to data base accessible
for subsequent off-line analysis.

To define the cross section of a nuclear reaction, the
number of incident particles has been counted taking into
account the dead time. For 3 runs with a total duration of 20
days this number was estimated to be $N_o=5.6\times10^{10}$ while the
number of masters transferred into the computer was $N_m=1.18\times10^6$.

\section{\hspace{-0.8em}. Data processing}
\subsection{\hspace{-0.8em}. Preliminary data operation}
Preliminary data operation has been made on Pentium-Pro (OS Linux)
using MAGOFF code and included the reconstruction of the momentum of charged
particles registered by the spectrometer, the definition of photon energy
via the signal amplitude in the GS, the selection of events and the creation
of the data base needed for further analysis.

In the present work the hadron-inclusive events (with one charged particle
detected in the magnetic spectrometer) are analysed.  The search of candidates
was carried out among the events with simultaneous signals from at least 4
"y" planes  and 3 "z" planes placed in front of the magnet, 3 "Y"
and 2 "Z" planes behind the magnet. The particle momentum was defined after
the "front" and "back" tracks were joined by the line which
describes the trajectory in the relevant magnetic field.

Further selection of events was carried out according to the
following criteria. If the number of extra signals from any of the 4 groups
(y,z,Y,Z) exceeded 3, the event was rejected. It was shown by Monte-Carlo
calculations that this algorithm permits to minimize the
probable contribution of reactions with greater multiplicity as well as the
contribution of the background events.

A total of $0.242\times10^6$ one-track events were selected from
$1.18\times10^6$ masters.
The information send to the data base included seven parameters
for each 1-track event:
$X_{ti},Y_{ti},Z_{ti}$- the measured coordinates of beam interaction point
with the target nucleus, $E_{\gamma i}$ - photon energy,
$p_i,\theta_i,\pi_i$ - the momentum of secondary particle and the angle after
the interaction.

\begin{figure}[hbt]
\begin{minipage}{0.45\linewidth}
\begin{center}
\mbox{ \epsfysize=5.5cm \epsffile{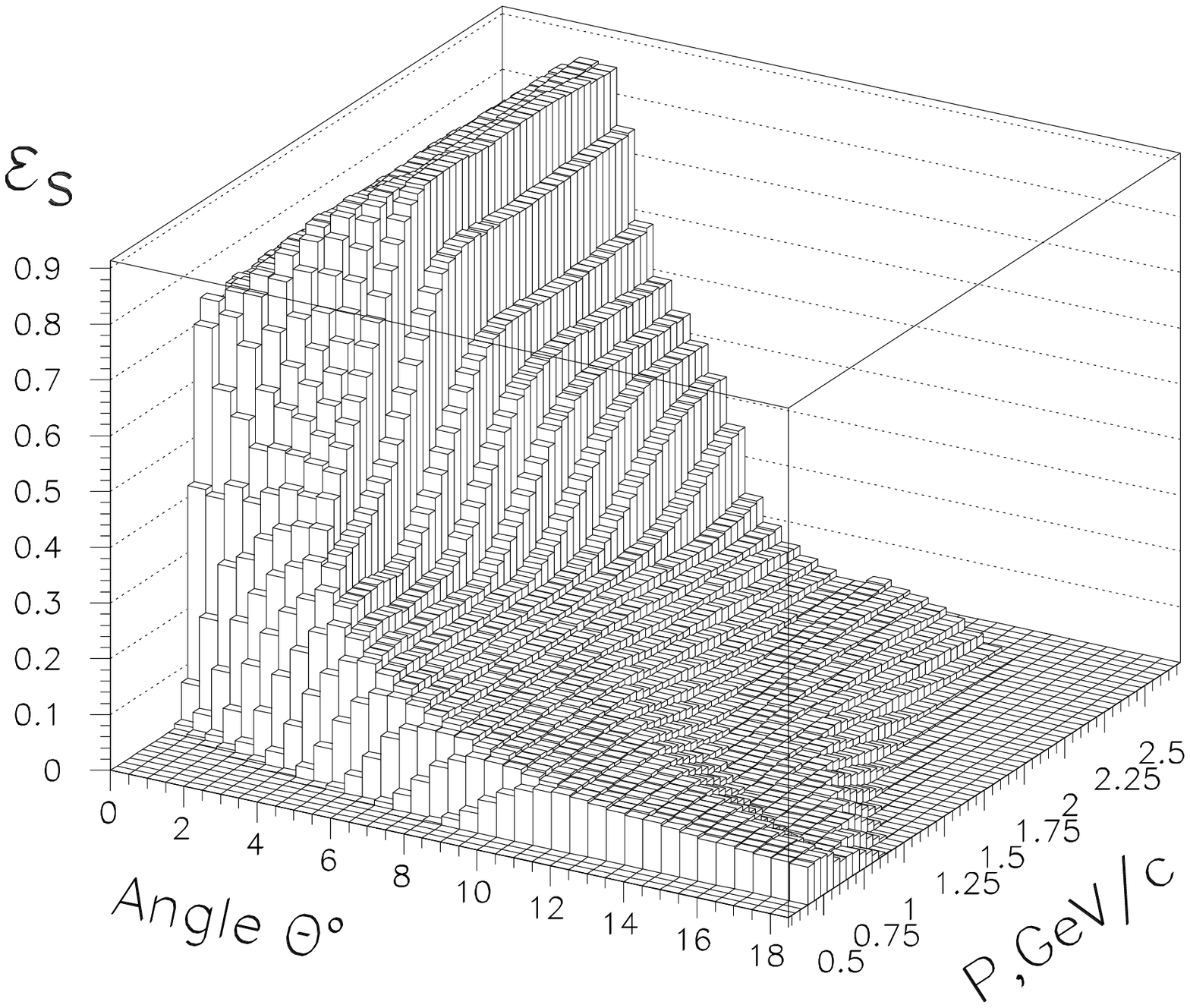} }
\end{center}
\caption{MAGE registration efficiency $\varepsilon_{s}$
as a function of the angle and momentum.}
\label{mageps1}

\end{minipage}
\vspace{-8.5cm}

\begin{flushright}
\begin{minipage}{0.45\linewidth}
\mbox{ \epsfysize=5.5cm \epsffile{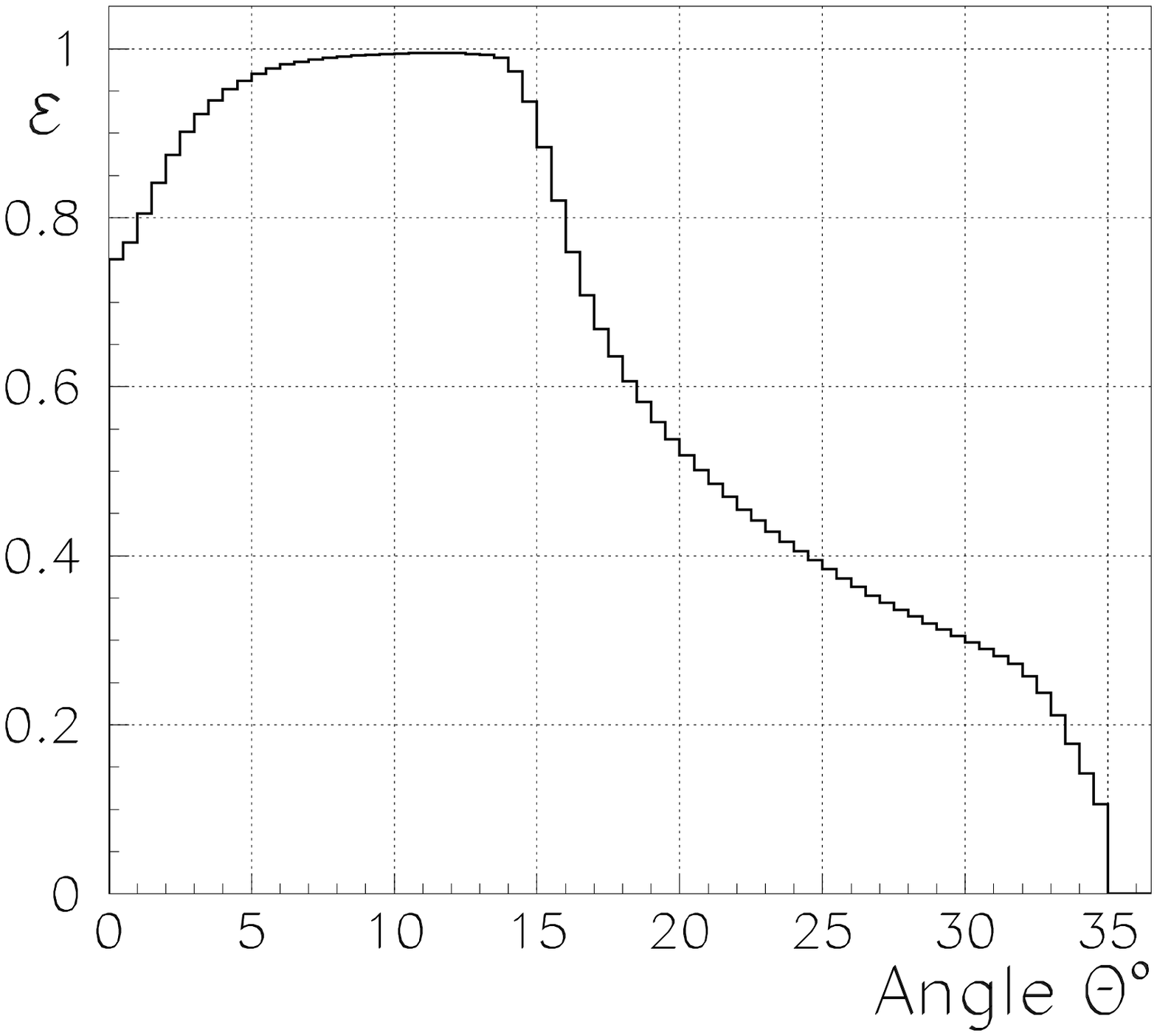} }
\caption{The efficiency $\varepsilon$ of particle detection in
forward blocks PC1 and PC2 as a function of the particle angle.}
\label{mageps2}
\end{minipage}
\end{flushright}
\end{figure}

\subsection{\hspace{-0.8em}. The efficiency of the magnetic spectrometer}

The efficiency of the magnetic spectrometer includes the geometric
efficiency as well as the chamber efficiency $W$ calculated for
each working proportional chamber using MAGOFF code.
For a number of chambers it was
found to be dependent on the hit position:  $W=U(y,z)$.
 This information was used in Monte-Carlo (GEANT) simulation
of the spectrometer efficiency $\varepsilon_{s}=f(\theta,p)$ as function of
the angle and momentum of the particle emitted from the target.
  The efficiency $\varepsilon=f(\theta)$ of particle registration by the
  forward blocks PC1 and PC2 has been also calculated.
All the calculations were made for the uniform azimuthal distribution of
secondary particles. The results are shown in fig.4 and fig.5.

 The calculated efficiencies were taken into account in experimental data
 analysis where every i-th event was weighted by
$b_i(\theta_i,p_i)=1/\varepsilon_{si}(\theta_i,p_i)$.

\subsection {\hspace {-0.8em}. The excitation energy resolution
of the spectrometer}

The nucleus excitation energy $\omega$ is one of the main parameters
obtained for quasi-coherent scattering on nuclei.
In the case of one leading particle it can be defined as the difference
between the missing mass $M^*$ in the reaction

\begin{equation}
\label{freac}
p~+~^{28}Si~\rightarrow~p^{\prime}~+~^{28}Si^*
\end{equation}

and the rest mass $M$ of the target nucleus:

\begin{equation}
\label{fomega}
\omega=M^*-M
\end{equation}

The missing mass $M^*$ can be expressed as

\begin{equation}
 M^* =\sqrt{(E_0+M-E)^2-p_0^2-p^2+2p_0p~cos\theta} ,
\end{equation}

where ($E_0,p_0$) and ($E,p$) - is the energy and the momentum of
incoming and registered protons respectively.
To analyze $\omega$ spectra it is necessary to know the MAGE excitation
energy resolution function $g(\omega)$ which
depends on the momentum resolution of the spectrometer as well as on
the momentum dispersion of beam particles.
The shape of the function (see Fig.6) was obtained from $\omega$
distribution with excitation energy $E^* = 0$
approximated by a sum of two Gaussian functions (smooth line):

\begin{figure}[hbt]

\begin{center}

\mbox{ \epsfysize=9.5cm \epsffile{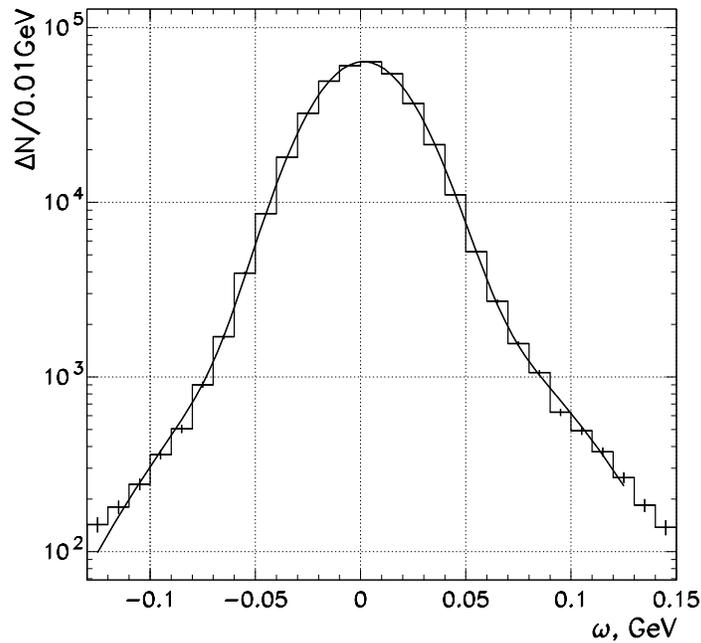} }

\end{center}

\caption{The excitation energy resolution function $g(0,\omega)$.}

\label{resol}

\end{figure}

\begin{eqnarray}
\label{fresol}
 g(E^*,\omega)=exp(-0.5 ((\omega - E^*)/(0.0225\pm 0.0005))^2) + \hspace{3cm}
\qquad \nonumber\\
 (0.075\pm0.001)~exp(-0.5 ((\omega - E^* +
0.0085\pm0.0003)/(0.0500\pm0.0007))^2).
\end{eqnarray}
where $E^*$ is the energy of the observed excited level expressed in GeV.

The first term of this formula is responsible for 93\% contribution and
 has a standard deviation of 0.0225 GeV.
In the first approximation this value can be treated as the
energy resolution of the magnetic spectrometer.

\section{\hspace{-0.8em}. Measurement of the cross section of the reaction
$^{28}Si(p,px)^{28}Si^*$ $(E_{\gamma}=1.78~\mbox{MeV})$}

Reactions of quasi-coherent interaction with $^{28}Si$ nuclei
have been selected via 1.78 MeV characteristic
$\gamma$ radiation following nuclear transition from first
excited state $2^+_1$ into the ground state $0^+_{gs}$.
The choice of this level was motivated by the results of Monte-Carlo
calculations. It was shown that the braking time of Si nucleus
after the interaction with a proton is far less then the half-decay period
and the shape of the $\gamma$-line is not distorted by Doppler effect.

The $\gamma$ energy essential for the selection of reactions was
calculated for all $1.18\times10^6$ registered events via the
amplitude $A_{gd}$ with the help of the calibration dependence (1).
The photon energy distribution for the region
1.73$\div$1.85 MeV is shown in Fig.7.

\begin{figure}[hbt]

\begin{center}

\mbox{ \epsfysize=9.5cm \epsffile{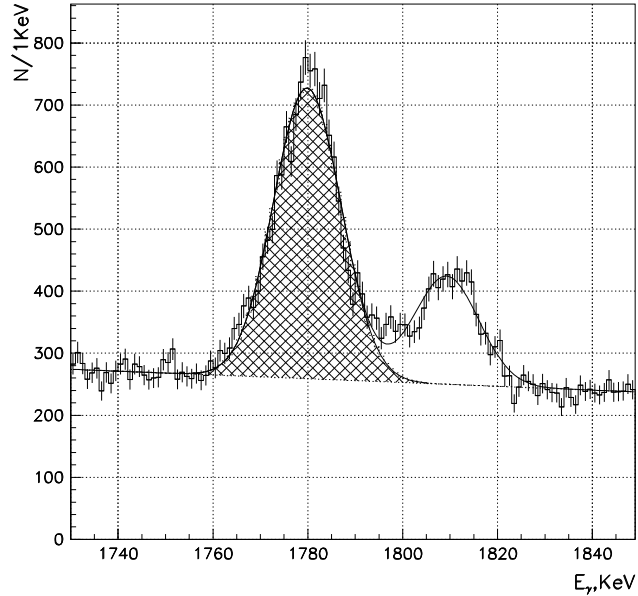} }

\end{center}

\caption{ $\gamma$ energy distribution in
$^{28}Si^*(2^+_1 \rightarrow 0^+_{gs})$ transition region.}

\label{egtot}

\end{figure}

To the right side of the analyzed peak a background signal of reaction
$^{28}Si(p,2p)^{26}Mg^*(2^+_1,1.809 MeV)$ with subsequent transition of
$^{26}Mg^*$ nucleus into the ground state can be seen.

The distribution was approximated by a sum
of a liner polinom (background) and two Gaussian functions. The solution
found by the method of least squares is shown by smooth lines in Fig.7.

For the analysed peak the mean value of $\gamma$ energy was found to be
$E_\gamma=(1.7799~\pm~0.0002(stat.)~\pm~0.0010(syst.)$) MeV
(which is in good agreement with the tabulated value ~\cite{Endt}) and
the standard deviation - $\sigma_{E_{\gamma}}$= 7.1 keV.
The number of quasi-coherent events was calculated as the area under the peak
(hatched) equal to $8656 \pm 161$  less background from the reactions
$^{29}Si(p,pn)^{28}Si^* and ~~^{30}Si(p,p2n)^{28}Si^*$  equal to 50$\pm$10.
It is necessary also to take into account the probable contribution from
the process of $^{28}Si$ disintegration by protons in which the
de-excitation of the residual nucleus is followed by photon emission
in the area $E_\gamma\pm2\sigma_{E_{\gamma}}$:

$^{25}Al(7/2^+,2.72~\longrightarrow\hspace{-1.cm}^{1.775}~3/2^+,0.944)$,~~
$^{25}Al(5/2^+,1.79~\longrightarrow\hspace{-1.cm}^{1.790}~5/2^+,gs)$,\\
$^{26}Mg(0^+,3.589~\longrightarrow\hspace{-1.cm}^{1.780}~2^+,1.809)$,~~
$^{26}Mg(4^+,5.715~\longrightarrow\hspace{-1.cm}^{1.775}~3^+,3.941)$.

Since no experimental data are available on the probabilities of
relevant nuclear reactions, in order to estimate the contribution of
these reactions the alternative channels of de-excitation with
photon emission in the energy range of $50~\div~3500$ have
been used. The upper limit for the contribution was
estimated to be 250 events with 67\% confidence level for 4 mentioned
$\gamma$-transitions.  Taking into account these corrections we obtain
$N_{1.78}=8606^{+161}_{-300}$ events.

The cross section of quasi-coherent interaction with
the discharge of excited states through 1.78 MeV level has been determined
as:

\begin{equation}
\sigma_{1.78}=\frac{4\pi~A~N_{1.78}}{N_{av}~\rho~x~\Omega_{eff}(E_\gamma)~k~\xi~
N_0}, = (37.8^{+0.8}_{-1.5}(stat.)~\pm~4.0~(syst.))\mbox{mb},
\end{equation}
where A=28 is the atomic number of the target nucleus;
$N_{av}$ - the Avogadro number,
$\rho$=2.20~g/cm$^3$ and $x$=2.8~cm - the density and the thickness of
the target;
$\Omega_{eff}(E_\gamma)$ - Ge detector efficiency; $k$=0.65 -
$\gamma$ absorbtion coefficient (calculated by the Monte-Carlo method);
$\xi$ = 1.05 - the coefficient of $gamma$ angle distribution
in nuclear transitions $2^+\rightarrow0^+$ ~\cite{Kirp4};
$N_0=5.6\times10^10$ - the number of beam particles reaching the target.

The main peculiarity of first excited level $^{28}Si^*(2^+,1.78)$ is the
probable "contribution from above", since at least 10 excited
levels with the energy less then the nucleon knock-out energy $E_t$ have a
considerable probability of de-excitation through this level
(~\cite{Endt},~\cite{Vergam}).
Therefore we don't know exactly which excited state resulted from
the interaction with protons.  However, the comparison of our result
with other experimental data allows to make a number of estimations which
are presented in the next section.

\section{\hspace{-0.8em}. The analysis of hadron inclusive quasi-coherent events}

\subsection{\hspace{-0.8em}. The selection of events}

 A total of 252163 events with one charged particle (proton mainly) registered
  in the magnetic spectrometer were selected in all $\gamma$ energy range.

 It was shown by Monte-Carlo
calculations that small ($<2\%$) admixture of $\pi^{+}$-mesons from
$\pi^{+}\pi^{-}$ pair production can be present among the registered
particles.  In order to reject the events without the
beam proton interaction with the target, the restriction was imposed on
the proton scattering angle with regard to the beam direction:
$\theta~>$~1.5$^0$

For the study of $^{28}Si(p,px)^{28}Si^*$ quasi-coherent reactions events
accompanied by 1.78 KeV photon emission which correspond to
$^{28}Si^*(2^+_1 \rightarrow 0^+_{gs})$ transition were selected.
  The photon energy spectrum in the transition area is shown in fig. 8.
It is necessary to note once more that all the distributions are
efficiency corrected (see section 3.2).

The area 1760$\le$~E$_\gamma~\le$~1790 keV chosen for the
selection of quasi-coherent events, is marked by dark hatching.
To estimate the background contribution, the distribution
on the left and right sides from the maximum was approximated by a line.
It was obtained that the number of events in the selected interval is equal
to 2380, the background - 560$\pm$~50, and the number of quasi-coherent
events - 1820~$\pm$~50. The events on the right and left side from the peak
(marked by rare hatching) were used for the establishment of
the shape of the background distribution of the nucleus excitation energy
and of the other kinematic variables.

\begin{figure}[hbt]

\begin{center}

\mbox{ \epsfysize=9.5cm \epsffile{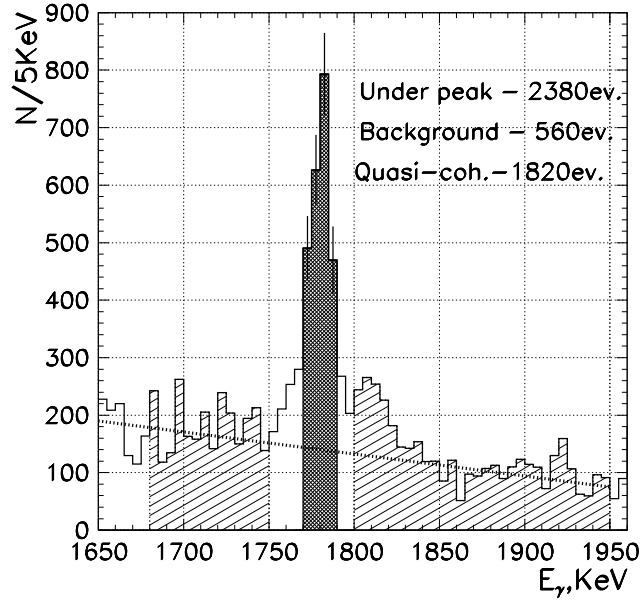} }

\end{center}

\caption{$\gamma$ energy distribution of one-track events in
$^{28}Si^*(2^+_1 \rightarrow 0^+_{gs})$ transition area.}

\label{egtot}

\end{figure}

The same calculations have been performed for the events with one
charged particle detected in the forward blocks of the proportional chambers
PC1 and PC2. Since the angle acceptance was better in this case,
a larger number of quasi-coherent events was registered:
2701~$\pm$~65 with the background of 1435~$\pm$~45.

\subsection{\hspace{-0.8em}. Differential cross sections}

Differential cross sections have been obtained taking into account
three corrections.

1. The efficiency of the selection of one-track events by the MAGOFF code
(0.61~$\pm$~0.05). It was estimated via the beam expositions (only
one-track events by definition).

2. The missing events with proton scattering at $\theta>$
30$^\circ$ . The share of such events was estimated considering the
number of events without the response in the proportional chambers.
It was found to be 22\% from the total number of registered events.

3. The admixture of events with the larger multiplicity of charged
particles ($<1$\%).

Fig.9 shows the differential cross section distribution of 2701 events
registered without the momentum measurement (solid line) as well as the
distribution of 1820 events with the momentum of $\ge$ 0.4 GeV/c
(dotted line). Both distributions are background subtracted.
Their agreement within the statistical errors in the angle range up to
$15^\circ$ points to the absence of noticeable systematic
errors in MAGE efficiency.

\begin{figure}[hbt]

\begin{center}

\mbox{ \epsfysize=9.5cm \epsffile{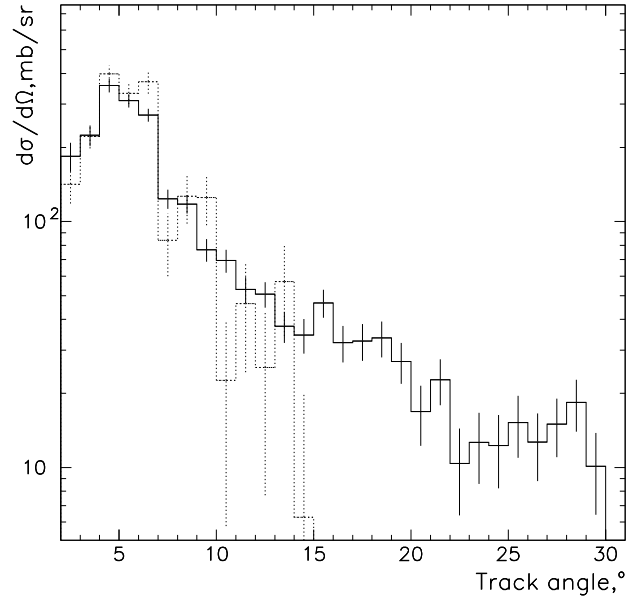} }

\end{center}

\caption{Differential cross section of the reaction
$^{28}Si(p,px)^{28}Si^*(E_{\gamma}=1.78~\mbox{MeV})$ for events detected in
the forward chamber blocks PC1 and PC2 (solid histogram) and for events,
registered in all blocks of chambers (dotted histogram).}

\label{dsigdom1}

\end{figure}

It is interesting to compare the obtained cross sections with the results of
other works.  The differential cross sections of 1 GeV proton quasi-coherent
scattering on $^{28}Si$ followed by the excitation of first three levels were
measured by LINF group in 1975-1979 at the magnetic spectrometer with the
energy resolution of 2 MeV ( see ~\cite{Alhaz3},~\cite{Alhaz4}).
The integration of this cross section over the proton scattering angles
in the interval 5.7--15$^\circ$ results in the value 8.9~$\pm$~0.9 mb for
levels 1.78, 4.61 and 4.96 MeV.  Since the transitions from 4.61 and 4.96
levels always proceed through the level 1.78 they are registered in
our experiment. Unfortunately, we are not able to separate these
reactions because of distortions of $\gamma$-spectrum by Doppler effect.
However, selecting the events with $\omega \le$ 0.05 GeV we can
estimate in the same angle range the excitation cross section for all
$^{28}Si$ levels with 1.78 $\gamma$
 transition as a final stage of de-excitation process.  It was found to be
$\sigma(5.7^\circ~\le~\theta~\le15^\circ,~\omega~\le~0.05~\mbox{GeV})=
(11.5~\pm~0.5~\pm~1.2)$~mb which agree within the limits of two standard
deviations with the LINF resutls. The differential cross-sections of the
reaction $^{28}Si(p,px)^{28}Si^*(E_{\gamma}$=1.78~MeV) obtained in our
experiment and LINF data are shown in fig.10.

\begin{figure}[hbt]

\begin{center}

\mbox{ \epsfysize=9.5cm \epsffile{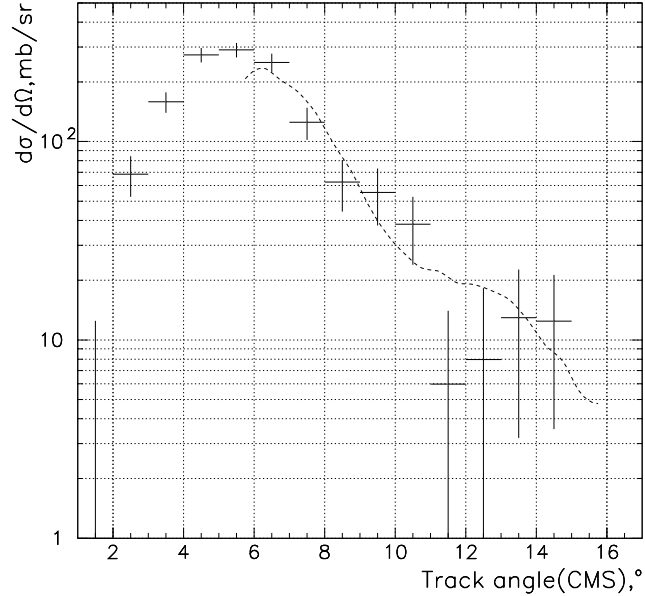} }

\end{center}

\caption{Differential cross section of the
$^{28}Si(p,px)^{28}Si^*(E_{\gamma}$=1.78~MeV) reaction in the center of mass
system for the events with $\omega<$~0.05~GeV. LINF result is shown by
smooth line.}

\label{dsigdom}

\end{figure}

However, the cross-section obtained in our experiment is higher by
$\approx$25\% which can be interpreted as an indication of excitation
of $^{28}Si$ levels with the energy greater than 4.96 MeV
in quasi-coherent interactions.
The cross section of such processes can be estimated as
the difference of two experimental results:

\begin{center}
$\sigma(5.7^\circ~\le~\theta~\le15^\circ,E^*~\ge~4.96)=2.6~\pm~1.2~\pm~1.2$ ö.
\end{center}

\subsection{\hspace{-0.8em}. Analysis of the excitation energy
spectrum. Search for highly excited states}

The energy transfer $\omega$ distribution of 2380 candidates for
quasi-coherent events selected in the interval
1760 keV $\le$~E$_\gamma~\le$~1790 keV is shown in fig.11
in logarithmic scale (dotted histogram).

\begin{figure}[hbt]

\begin{center}

\mbox{ \epsfysize=9.5cm \epsffile{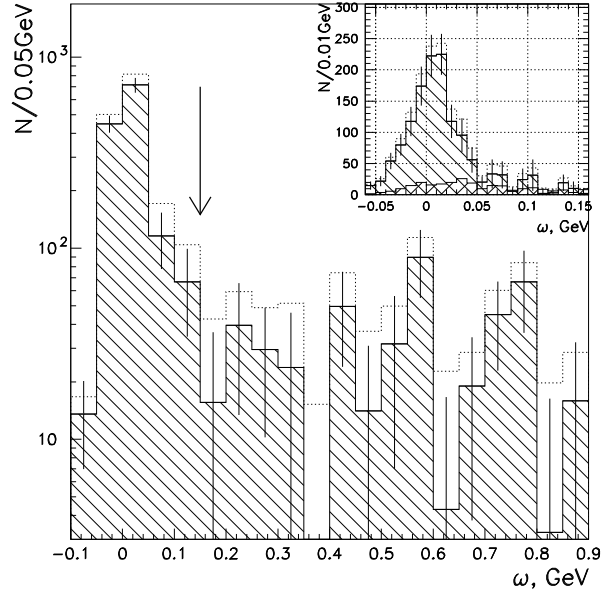} }

\end{center}

\caption{ $\omega$ distribution for 1760~keV~$\le$~E$_\gamma~\le$~1790~keV.
Hatched is the background subtracted distribution (1820 events).
The regions of elastic scattering (left) and unelastic interactions  are
separated by arrow. In right top corner - $\omega$ distribution for
elastic events with the background (560 events) shown by double hatching.}

\label{h59}

\end{figure}

1820 events left after the background subtraction are shown by the hatched
histogram. A bulk of events can be seen in the region of small excitation
energies $\omega$, at larger energies the distribution is more uniform.
The arrow in fig.11 indicates the value $\omega$=0.15 GeV separating two
energy regions: the region of non-meson excitation or elastic interactions
on the left (1369 events) and the region of possible $\pi$ - meson production
on the right (461 events).

\subsubsection{\hspace{-1em}. The region of quasi-coherent elastic
interactions ($\omega \le$~0.15~GeV).}

It was shown above that this region is dominated by the processes of
excitation of the first three $^{28}Si$ levels with the average energy
$\overline{E^*_{1-3}}=3.3~$MeV. Since the gap between the levels
(3.22 MeV) is far less then the
MAGE energy resolution ($\approx$~22 MeV), the spectrometer response
must have the same shape as the excitation energy resolution function
$g(\overline{E^*_{1-3}},\omega)$ (5). To take into account the possible
contribution from the excitation of higher energy levels $\omega$
experimental spectrum has been approximated by the sum of three
resolution functions:

$F=a_1\cdot~g(\overline{E^*_{1- 3}},\omega) + a_2\cdot~g(a_3,\omega) +
a_4\cdot~g(a_5,\omega)$

 where the parameters $a_3$, $a_5$ are responsible for the
energies of levels and $a_1$, $a_2$, $a_4$ - for their possible contributions.
The values of parameters were calculated by the maximum likelehood method:
$a_1$=147~$\pm$~46, $a_2$=50~$\pm$~45,
$a_3$=0.013~$\pm$~0.010, $a_4$=17~$\pm$~8, $a_5$=0.099~$\pm$~0.006.
This solution is presented in fig.12, where the function $F$ (solid line)
and three components marked by figures are superimposed on the
experimental histogram.

\begin{figure}[hbt]

\begin{center}

\mbox{ \epsfysize=9.5cm \epsffile{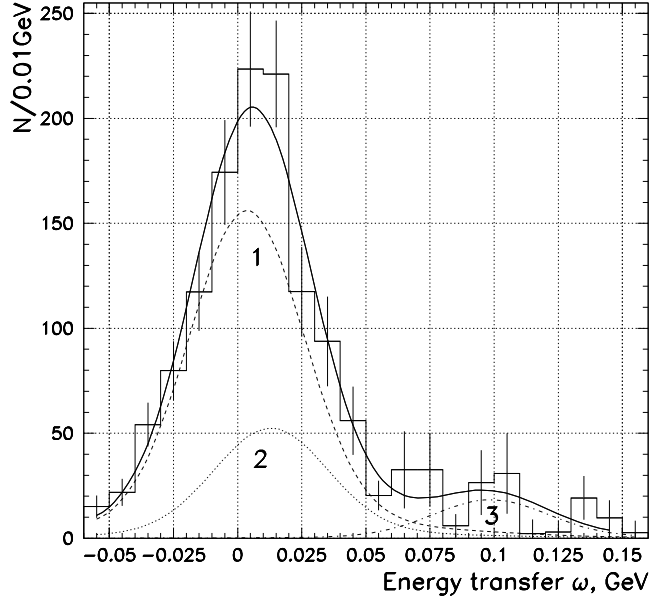} }

\end{center}

\caption{The results of $\omega$ distribution fit by function $F$.}

\label{omcoha}

\end{figure}

Except the processes of excitation of first three levels (curve 1), which
contribute approximately 70\%, the minimization code finds two
non-zero contributions in 10 MeV (curve 2) and
100 MeV (curve 3) regions. The contribution from levels with
4.96~MeV$\le~E^*~\le~E_t$ seems probable enough and don't
contradict the conclusion made in section 5.2.
In order to test the probability of the existence of nuclear level
described by the third component of function $F$, the comparative analysis
of $\gamma$ specrta for the events with lower ($\omega~<~$0.06~GeV) and
higher (0.06~GeV~$<~\omega~<~$0.15~GeV) excitation energies has been made.
(see fig. 13).

\begin{figure}[hbt]

\begin{center}

\mbox{ \epsfysize=9.5cm \epsffile{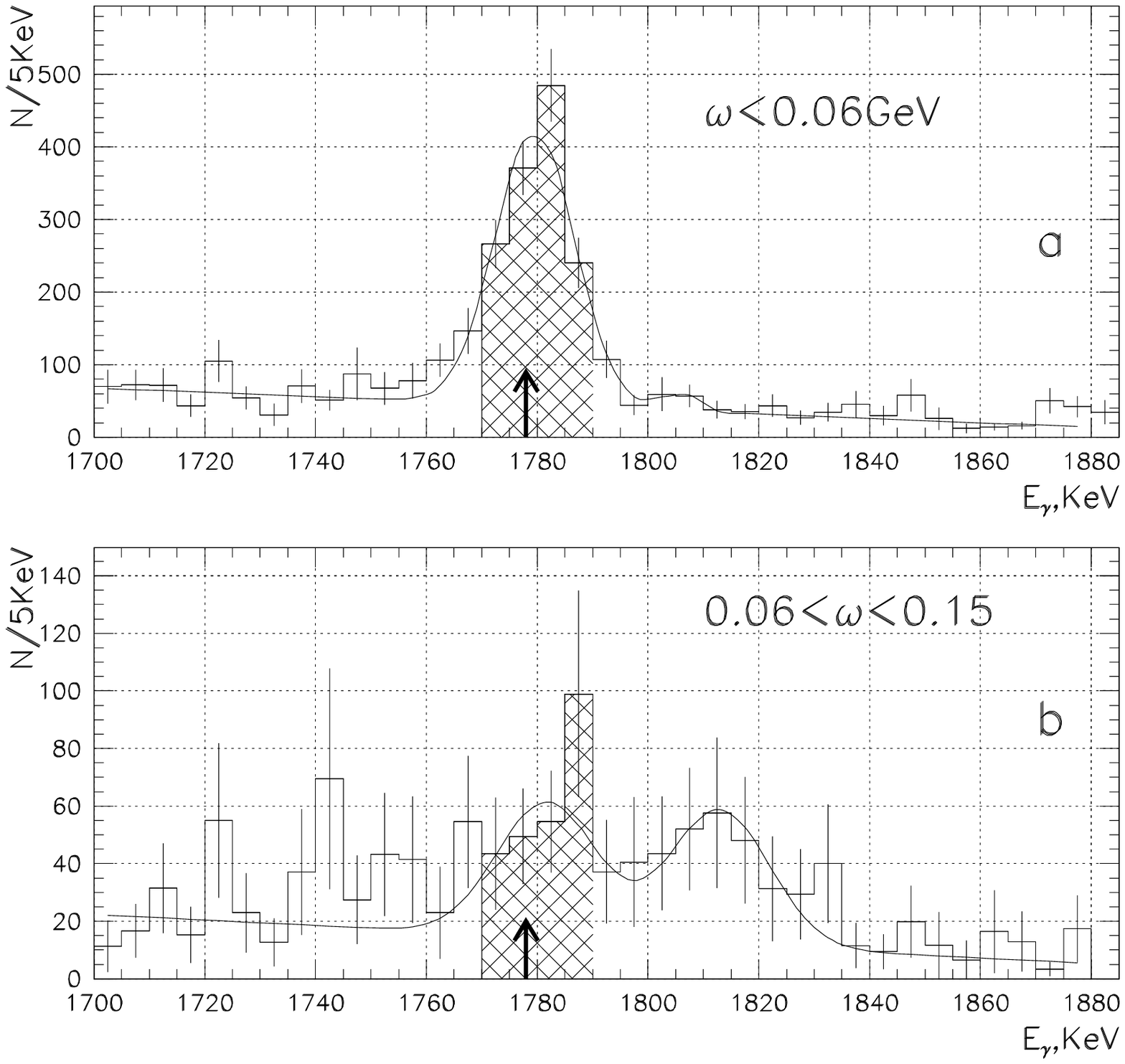} }

\end{center}

\caption{$\gamma$ energy spectra in the transition region
$^{28}Si^*(2^+_1 \rightarrow 0^+_{gs})$  for two excitation energy intervals:
$\omega$ $\le$~0.06~GeV(a) and 0.06~$<~\omega~<$~0.15~GeV(b).
The energy of the 1.78~MeV transition is indicated by arrows.}

\label{fitg1h2}

\end{figure}

The selected quasi-coherent events (see section 5.1) displayed in fig.12
as a histogram are marked by hatching. The level of 1.78 MeV is indicated
by arrows. The clear maximum is observed for quasi-coherent reactions
at small energies (fig.13a). The absence of a noticeable signal in the hatched
area in fig.13b indicate that the events observed in the region
$\omega$=0.06~$\div$~0.15 GeV could be the result of a background fluctuation
and can be used only to set up the limit on the excitation cross section
of long lived nucleus levels. The calculation results are presented
in section 5.3.3.

Supposing that the level described by the third component of the function $F$
is imitated by the background fluctuation, in order to determine the
average excitation energy of $^{28}Si$ nucleus following quasi-coherent
elastic interactions with protons $\overline{E^*}$ the experimental
distribution in the region $\omega~\le$~0.06 GeV has been described by the
resolution function $F_1=b\cdot~g(\overline{E^*},\omega)$ only.
The estimation of $\overline{E^*}$
by the least squares method gives $\overline{E^*}$=0.0053~$\pm$~0.0014~GeV
with $\chi^2$ equal to 6.9 for 9 degrees of freedom.

\subsubsection{\hspace{-1em}. The region of $\pi$-meson production
($\omega~\ge$~0.15~GeV).}

In the region $\omega~\ge$~0.15 GeV the proton transfer energy is sufficient
for $\pi$-meson production.
It is known that at proton energies of about 1 GeV $\pi$-mesons
are produced on free nucleons mainly via the isobars
$P_{33}(1232)(\Delta)$ and $P_{11}(1440)$ (~\cite{Chiba},
~\cite{Niita},~\cite{Verwest}). Suggesting the same mechanism for
$\pi$-meson production on quasi-free nucleons in quasi-coherent proton
interactions with nuclei, let's estimate the cross-sections.
 It is shown in ~\cite{Bianchi1}, ~\cite{Kondr}, ~\cite{Bianchi2} that the
production of $P_{11}(1440)$ is strongly suppressed and for the production
of $D_{13}(1520)$ the energy of the incoming proton apparently is not
sufficient since the upper limit for the X mass in the
reaction $pp \rightarrow pX$ at E=1 GeV is equal to 1.43 GeV/c$^2$.
Therefore the processes of one $\pi$-meson production were simulated using
$\Delta$ isobar only and the processes of two $\pi$-mesons
production - without the participation of isobars.

To estimate the cross section of $\pi$ - meson production
the experimental spectrum has been fitted
to a sum of $\omega$ distribtions obtained by Monte-Carlo technique for
three processes:

1. Nuclear excitation to the energy $\overline{E^*}$=0.053 GeV in the
reaction $^{28}Si(p,p^{\prime})^{28}Si^*$.

2. The isobar production on intranuclear nucleon $p_f$ ($n_f$) and the
decay through well known channels with $\pi$-meson production. Seven
reactions have been simulated:

\setcounter{equation}{0}

\begin{eqnarray}
\label{fdel1}
p~+~n_f~\rightarrow~n~+~\Delta_{1232}^{++}~(p\pi^+),~ ~    [0.47] \\
p~+~p_f~\rightarrow~p~+~\Delta_{1232}^+~(p\pi^o~),~ ~ ~    [0.14] \\
p~+~p_f~\rightarrow~p~+~\Delta_{1232}^+~(n\pi^+), ~ ~     [0.07] \\
p~+~n_f~\rightarrow~n~+~\Delta_{1232}^+~(p\pi^o~),~ ~     [0.11] \\
p~+~n_f~\rightarrow~n~+~\Delta_{1232}^+~(n\pi^+), ~ ~    [0.05] \\
p~+~n_f~\rightarrow~p~+~\Delta_{1232}^o~(p\pi^-), ~ ~    [0.05] \\
p~+~n_f~\rightarrow~p~+~\Delta_{1232}^o~(n\pi^0). ~ ~    [0.11]
\label{fdel2}
\end{eqnarray}

3. The production of a $\pi$-meson pair with the total charge equal to zero
on intranuclear nucleon in the following reactions:

\begin{eqnarray}
\label{f2pi1}
p~+~p_f~\rightarrow~p~+~p~+2\pi, \\
p~+~n_f~\rightarrow~p~+~n~+2\pi .
\label{f2pi2}
\end{eqnarray}

For the simulation of the last 2 processes the Fermi-momentum of the
target-nucleon was taken into account. According to the conservation law
the total energy of the target-nucleon was accepted to be equal to the
rest mass of a free nucleon.

For the reactions (1)-(7) the angle distribution in the center of mass system
was simulated according to the results of ~\cite{Niita}.  The contribution of
each reaction obtained from the experimental data on cross section of
$\pi$-meson production in different isotopic conditions of $\pi$N system
(see ~\cite{Niita},~\cite{Verwest}, ~\cite{Chiba}) is given in brackets.
The relative contributions of reactions (8) and (9) were accepted to be equal
since their simulated $\omega$ spectra have close shapes.
Quasi-coherent events have been selected according to the condition that
the nucleon energy after the interaction must be less then the energy of
nucleon splitting from $^{28}Si$
(0.0116 GeV for proton and 0.0172 for neutron).

\begin{figure}[hbt]

\begin{center}

\mbox{ \epsfysize=9.5cm \epsffile{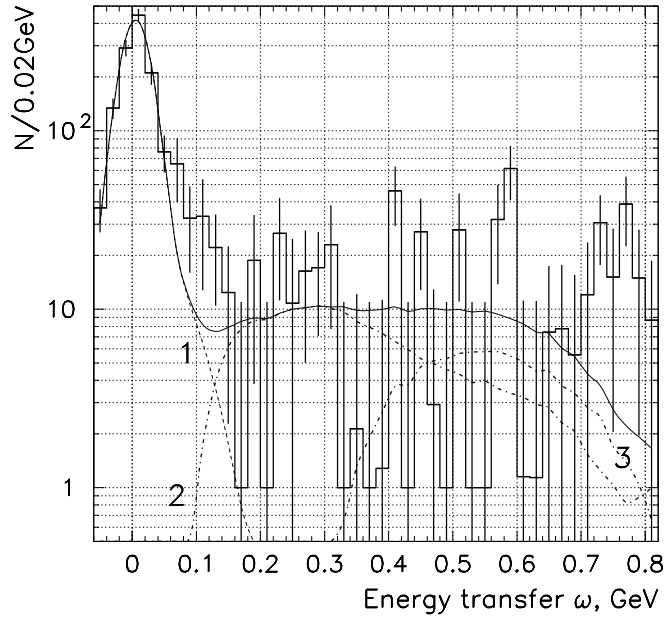} }

\end{center}

\caption{$\omega$ distribution of quasi-coherent events. Histogram - the
experimental data, solid line - the result of data fit by three
processes: the nucleus excitation to the energy $\overline{E^*}$=0.053 GeV
(curve 1), the isobar production on intranuclear nucleon (curve 2) and
$\pi$-meson pair production (curve 3).}

\label{omcoh4}

\end{figure}

The fit was performed without normalization to the total number of events
in the experimental histogram. The results are presented in fig.14.
The contributions of 3 processes were found to be
1290$\pm$70, 198$\pm$70 and 92$\pm$60 respectively. Taking into account
the share of events where charged particles are out of the
registration region of the spectrometer we obtain the following
cross-sections:

\begin{eqnarray}
\sigma_1= (14.2~\pm~3.0~(stat.)\pm~1.5(syst.)) \mbox{ö}, \nonumber \\
\sigma_2= (7.5~\pm~3.0~(stat.)\pm~0.8(syst.)) \mbox{ö}, \nonumber  \\
\sigma_3= (3.1~\pm~2.0(stat.)~\pm~0.3(syst.)) \mbox{ö}.  \nonumber
\end{eqnarray}

It should be noted that  $\sigma_1+\sigma_2+\sigma_3=24.8 mb$
provides only the contribution of 2/3 to the total cross section of
quasi-coherent scattering (37.8 mb) which indicates the
probability of other quasi-coherent processes, in particular,
the meson resonance production on a nucleus.

It is interesting to compare the two latest values to the cross sections
of the same reactions on free nucleons. As for the quasi-coherent interaction
$\sigma_3/\sigma_2~\approx~40~\%$, for free nucleons this relation is
$< 10$\%.
 We can't give valuable explanation to the fact up to now,
however the increase in the production of two-pion system following the
hadron interactions with nucleus has been observed by a number of recent
experiments (see ~\cite{2pi1},~\cite{2pi2}).
The mechanism of this phenomenon is not established yet.

\subsubsection{\hspace{-1em}. Limitation on the excitation cross section
of high energy levels}


It is seen from fig.14 that in several regions of energy transfer
the experimental spectrum is not well described with the
selected combination of the simulated distributions.

\begin{figure}[htb]

\begin{center}

\mbox{ \epsfysize=9.5cm \epsffile{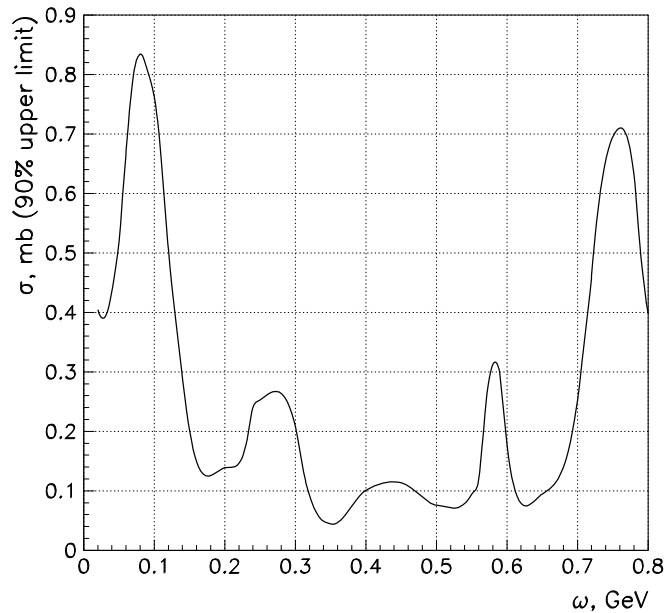} }

\end{center}

\caption{The upper limit of the excitation cross section of high energy levels
in quasi-coherent interactions of 1 GeV protons with
$^{28}Si$ at different excitation energies.}

\label{lev90}

\end{figure}

\begin{figure}[ph]

\begin{center}

\mbox{ \epsfysize=15.5cm \epsffile{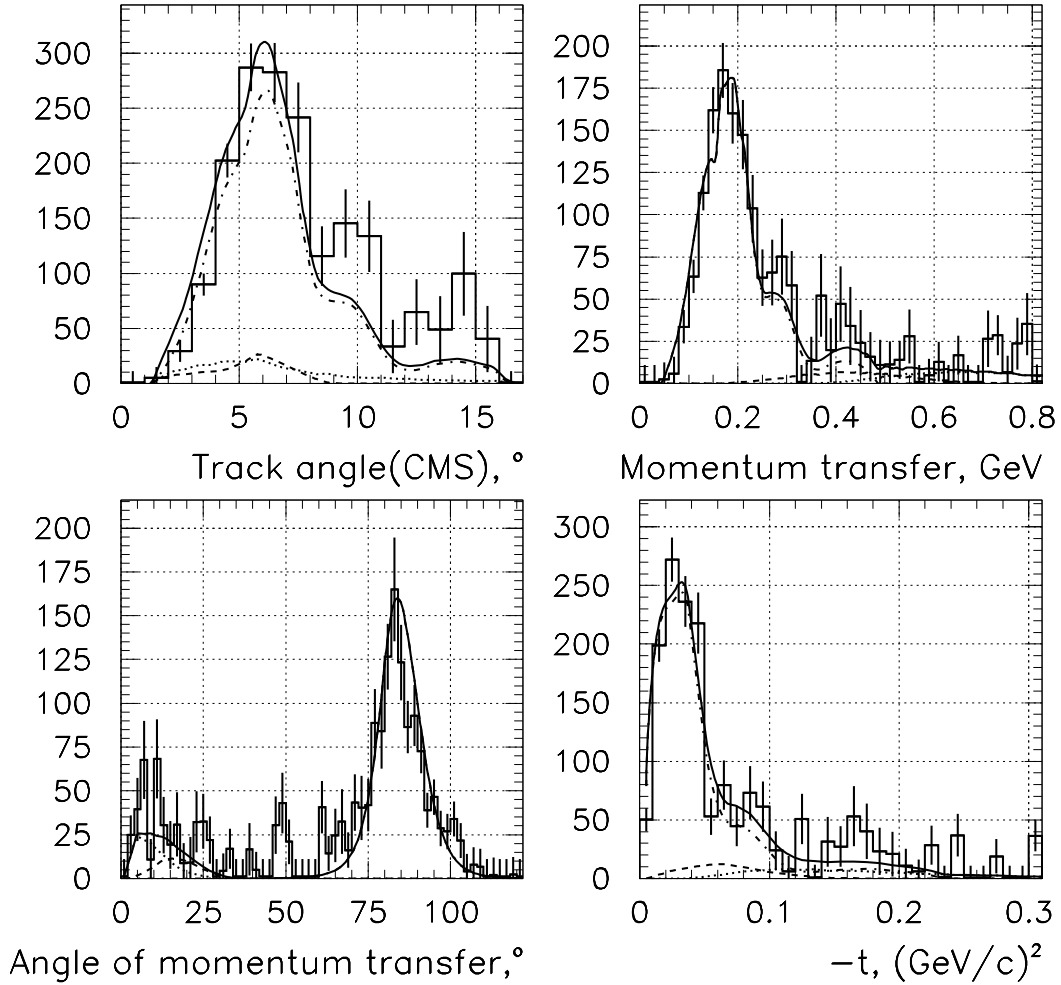} }

\end{center}

\caption{The distributions of a) the proton scattering angle in
the center-of-mass system, b) the momentum transfer,
c) the angle between the beam direction and the momentum transfer
in the laboratory system, and d) the square of transferred 4-th momentum.
The simulation results are shown by a smooth line for 3 processes:
dash-dot - elastic interaction; dash - $\Delta$-isobar production;
dot - the production of two $\pi$-mesons.
The sum of 3 processes is shown by the solid line.}

\label{kin4h}

\end{figure}

One of the regions ($0.06~\mbox{GeV}<~\omega~<~0.15~\mbox{GeV}$)
in which the excitation of high energy level is probably displayed
was mentioned in section 5.3.1. The lack of statistics does not allow
to confirm the existence of such levels, it is possible only to impose
limits on the probability of the level excitation.
The upper limit for this process calculated by the method of least
squares at the 90\% confidence level is presented in fig.15.
It is close to 0.1-0.2 mb in the region of 0.3-0.6 GeV and increases to 0.9 mb
in the region of 0.1 GeV which points to the excitation of a
high level of $^{28}Si$ (curve 3 in fig.12).  The increase of the
upper limit in the region of 0.76 GeV can be possibly explained
by $\pi$-meson resonance production ($\rho$ and $\omega$) on the nucleus.

\subsection{\hspace{-0.8em}. Other kinematical parameters}

The selected quasi-coherent events have been analysed taking into account
other kinematical parameters. 4 distributions for 1820 events are presented
in fig.16:

a) the proton scattering angle distribution in the p--$^{28}Si$ center of mass
system; b) the momentum transfer distribution;
c) the distribution of the angle between the beam direction and the
momentum transfer in the laboratory system and d) the square of transferred
4-th momentum distribution.  The simulation results for 3 processes
are shown by smooth lines:  dash-dot line - elastic
interaction, dash - $\Delta$-isobar production, dot - the production of
two $\pi$-mesons.  The sum of 3 processes is presented by solid line.
 The comparison of the experimental and simulated distributions in fig.14
and fig.16 apparently suggests that the observed quasi-coherent reactions
can be described by the sum of three processes in all the region of
kinematical variables permitted by kinematical laws.

\section{\hspace{-0.8em}. Conclusion}

 The main results are the following:

1. For the first time the cross sections of quasi-coherent proton interaction
with $^{28}Si$ have been measured via the registration of $\gamma$-transition
from the first excited nucleus state to the ground state. The values
obtained are:
37.8$^{+0.8}_{-1.5}$(stat.)~$\pm$~4.0~(syst.)mb for the total cross section,
14.2$~\pm$~3.0~(stat.)$\pm$~1.5(syst.)mb for elastic cross section and
23.6~$\pm$~3.5~(stat.)$\pm$~3.0(syst.)mb for non-elastic cross-section.
\hspace{0.5cm}

2. Assuming that single $\pi$-mesons are produced on intranuclear nucleons
via the isobar $\Delta$(1232) and two $\pi$-mesons are
produced according to the invariant phase volume, the cross sections of these
processes have been obtained: 7.5$~\pm$~3.0~(stat.)$\pm$~0.8(syst.)mb
and 3.$~\pm$~2.(stat.)~$\pm$~0.3(syst.)mb respectively.  The two processes
summarized give only half contribution to the non-elastic quasi-coherent
cross section.
\hspace{0.5cm}

3. The differential cross sections of quasi-coherent interaction
have been measured.
\hspace{0.5cm}

4. The upper limit on the cross section of the excitation of long-lived highly
excited levels in $^{28}Si$ for all the region of energies transferred by
proton has been estabilished.
\hspace{1cm}

The authors express deep gratitude to the ITEP accelerator operation crew for
the presented opportunity of the experiment realization.
The authors are thankfull to G.A.Leksin, V.B. Gavrilov, A.I. Golutvin,
M.G. Schepkin for useful and critical remarks.

\clearpage

\renewcommand{\refname}

\end{document}